\documentclass{article}
\usepackage{amsmath,bm,graphicx,xcolor}
\usepackage[style=authoryear,backend=bibtex]{biblatex}
\usepackage{amsfonts}
\usepackage[left=30mm,right=30mm,top=20mm,bottom=20mm]{geometry}
\begin{document}
{\centerline{\Huge \bf Quartet, higher order and near resonant}}
\centerline{  \bf \Huge{interactions in nonlinear wave equations}}
\vspace{2mm}
\centerline{\large  \bf Alex Owen$^1$, Roger Grimshaw$^1$, Beth Wingate$^1$}
\vspace{2mm}
\centerline{$^1$College of Engineering, Mathematics and Physical Sciences, University of Exeter,}
\centerline{North Park Road, 
Exeter EX4 4QF, UK}
\section*{Abstract}
Motivated by problems arising in geophysical fluid dynamics, we investigate resonant and near resonant wave interactions in nonlinear wave equations with quadratic nonlinearity, We place a special focus on interactions between slow wave modes, with zero frequency in the linear limit, and fast modes. These regularly occur in geophysical fluid systems with conserved potential vorticity or similar conserved quantities. A general multi-scale asymptotic expansion is used to show how the higher order nonlinear interaction coefficients are derived as a combination of the first order terms arising at the triad interaction level. From the general nth-order interaction coefficient we present a proof by induction how the limiting effect for particular combinations of slow and fast modes pushes their interactions to a slower timescale, and we show how this is linked to the form of the conserved quantities. We compare near resonant expansions with exact resonant expansions, and show how near-resonances allow higher order expansions to be reduced to just the most dominant contributions. These contributions then occur at one order higher in the expansion when compared to the analogous exact resonance expansion.

\section{Introduction}
Throughout this article we consider nonlinear wave equations of the form:
\begin{align}
\frac{\partial \bm{u}}{\partial t}+ \mathcal{L}(\bm{u})=-\mathcal{N}(\bm{u},\bm{u}), \label{nl_eq_int1}
\end{align}
here $\bm{u}(\bm{x},t)$ is a dependent variable representing the state vector of the system, with spatial coordinates in vector $\bm{x}$ and time coordinate $t$, $\mathcal{L}$ is a linear differential operator acting on the spatial part of $\bm{u}$,which generates a dispersion relation when transformed to spectral space. $\mathcal{N}$ is a bilinear differential operator representing the quadratic nonlinearity; we briefly discuss above quadratic order nonlinearity in section \ref{discussion}. In particular the systems are conservative, with a conserved energy, and a second conserved quantity, related to a Casimir of the system. For instance, this structure regularly appears in geophysical fluid dynamics in the form of vorticity/enstrophy conservation laws. Although we are motivated by geophysical fluid dynamics, the results are derived using quite general assumptions and can be applied to other physical systems which meet our explicit assumptions.\\
\\
If there is a pointwise conserved quantity, for example potential vorticity on geophysical fluids, we can write one of the linearised evolution equations as:
\begin{align}
\frac{\partial Q}{\partial t}=0 \label{Qconstraint2}
\end{align}
For the linearly conserved quantity $Q$. This directly translates to the dispersion relation $\omega Q=0$, found by taking fourier transforms in space, with wavenumber $\bm{k}$, and in time, where $\omega$ is the frequency of the wave. Thus if $Q$ is not zero, then there is a wave mode with a zero frequency $\omega=0$, that is, a slow mode that does not evolve in the linearised system. Then any other mode, must have $Q=0$ to satisfy this relation. For this reason the first type of mode is usually termed `slow' in the fluid literature, and the second 'fast'. There is a considerable literature on the possible nonlinear interaction of such fast and slow modes, and at the quadratic level with only interactions the outcome is often that there is no interaction, see \cite{warn1986statistical}, \cite{embid1996averaging},  and \cite{grimshaw2007nonlinear}. Here we re-visit this, examining the role of conservation laws, and especially going beyond the triad interaction level.\
\\
For geophysical fluids $Q$ has a related integrally conserved quantity, the enstrophy, which usually takes the form of the integral of the square of the potential vorticity. This constricts the form of the asymptotic expansion and by continuing to higher order and considering the near resonant version of the usual triad and quartet relations we will derive some general constraints valid at all orders of the asymptotic expansion. \\
\\
Throughout this article we use the familiar multiple time scale asymptotic method.  Within the geophysical fluid dynamics community this method developed in the 1960s following the work of \cite{phillips1960dynamics}, although it was preceded by entirely separate work in other fields, such as \cite{ziman1960electrons} and even \cite{peierls1929kinetischen}. The method has since been honed by various authors(see \cite{craik1988wave}), notably \cite{benney1967propagation} and later work such as \cite{embid1996averaging}.\\
\\
A central theme of this article is the role of near resonances in these expansions. The near resonances have been previously explored as isolated triads for example in \cite{bretherton1964resonant}, and more recently by  \cite{grimshaw2007nonlinear}. These authors focused on the evolution of the isolated triad, which obeys an integrable triplet of equations. \cite{alber1998geometric} showed that the phases of the three waves may follow a homoclinic orbit such that they do not cycle indefinitely as in a resonant triad. This contrasts to the amplitudes, which necessarily cycle due to energy conservation.\\
\\
Work examining the effect of near resonances within a full system has mostly followed two routes. One of these is the empirical route, for example in \cite{smith2005near}, and \cite{smith1999transfer}, in which numerical simulations are compared in the cases where exact and near resonances as well as all possible interactions. It is found that the near resonances are necessary for the interpretation of qualitative features of the simulations.
The other main line of enquiry has been in the field of wave turbulence, for example, see \cite{nazarenko2011wave}. Here one major problem is finding exact resonances on a finite grid, which can be reduced to a number theory problem, see \cite{kartashova1990partitioning}, \cite{kartashova1998wave}. When near resonances are considered, the number of permitted interactions is larger and resonant clusters can be formed, see \cite{kartashova2008resonant}. Here we consider a similar scenario, but in an infinite spectral space. But instead of separating the motion into isolated triads or clusters of interacting waves we consider the clustering effect to be a symptom of reordering resonant sets from higher orders of the expansion where the scaling of the interaction causes faster evolution of the dynamics than is found in a regular exact resonant expansion.\\
\\
To establish notation and record the relevant literature, in section \ref{tri_ord} we derive the regular triad order expansion, for exact or near resonances. In section \ref{quart_ord} the next order of interactions is derived. This shows how slaved modes from non-resonant interactions link to form the slower resonant quartets. It is then shown that by including the near resonances at triad order the fastest of the quartet interactions are shifted to the faster triad timescale, represented within the triad evolution. In section \ref{high_ord} the expansion is taken beyond quartet order, and a completely general form of the higher order interaction coefficients is derived. Again it is shown that the near resonant triads then contain the equivalent time evolution as any higher order interaction, whereever the interaction coefficients are sufficiently large that the motion can be considered to occur on the triad timescale. Section \ref{nrsec_ntet_split} discusses possibly unexpected behaviour when different parts of the quartets and higher order constructions can be split between different orders of the expansion. In section \ref{zero_mode_implic} the conservation laws are used to derive restrictions on interactions coefficients, and by using the generalised form given in section \ref{high_ord} these are simply proven for all orders of the expansion.

\section{Multiscale asymptotic expansion}
\label{tri_ord}
We consider nonlinear wave equations in the form (1), which here we express in the form used by \cite{embid1996averaging}:
\begin{align}
\frac{\partial \bm{u}}{\partial t}+ \frac{1}{\epsilon}\mathcal{L}(\bm{u})=-\mathcal{N}(\bm{u},\bm{u}). \label{nl_eq_int}
\end{align}
Here $\epsilon $ is a small parameter measuring the weak nonlinearity, and the time scale is the corresponding slow time. The fast timescale for the waves is then $t/epsilon $, see below. We have a degree of freedom in the definition of this bilinear operator and so for simplicity we define it to be symmetric in its arguments such that: 
\begin{align}
\mathcal{N}(\bm{a},\bm{b})=\frac{1}{2}\left(\mathcal{N}(\bm{a},\bm{b})+\mathcal{N}(\bm{b},\bm{a})\right). \label{nonlinex}
\end{align}
In addition to energy conservation, we assume that there is a second pointwise conserved quantity corresponding to a Casimir of the system, as discussed above. In the example of fluid systems this Casimir corresponds to the various `vorticity' conserved quantities, those of the Kelvin circulation theorem.

A system that is conservative when linearised will always permit a full basis of eigenfunctions. For simplicity in the rest of this paper we will always assume that the boundary conditions permit Fourier transforms, such that we can write the system in the eigenbasis as follows:

\begin{align}
\bm{u}=\sum_{\bm{k},\alpha} a_{\bm{k}}^{\alpha}\bm{r}_{\bm{k}}^{\alpha} e^{i \bm{k}\cdot \bm{x} }e^{i\omega_{\bm{k}}^{\alpha} t}, \label{four_bas}
\end{align}
where $\bm{k}$ is a wavenumber, and the superscript $\alpha$ defines a particular mode of the system. 
The information on the state of the system is contained in the amplitude variables $a_{\bm{k}}^{\alpha}$, and the eigenvectors $\bm{r}_{\bm{k}}^{\alpha}$ along with the exponential functions defines a given eigenfunction. $\omega$ is derived from the dispersion relation for the linearised system of equations.\\
\\
We next derive the standard weakly nonlinear expansion up to triad order for exact resonances. This is well detailed in many sources, for instance \cite{craik1988wave}, \cite{embid1996averaging}. We define a fast time scale $\tau={t}/{\epsilon}$ as in \cite{embid1996averaging} and slow timescales $t_n=\epsilon^n t$, $n=1,2,...$ we write $\bm{u}=\bm{u}(\bm{x},\tau,t_0,t_1,...)$ as a function of multiple time scales. Then (\ref{nl_eq_int}) becomes:
\begin{eqnarray}
\frac{1}{\epsilon}\left(\frac{\partial \bm{u}}{\partial \tau}+\mathcal{L}(\bm{u})\right)=-\left(\frac{\partial \bm{u}}{\partial t_0}+\mathcal{N}(\bm{u},\bm{u})\right). \label{non_loc}
\end{eqnarray}
We expand the variable $\bm{u}$ as follows:
\begin{align}
\bm{u}(\bm{x},\tau,t_0,t_1)=\bm{u}_0(\bm{x},\tau,t_0,t_1)+\epsilon\bm{u}_1(\bm{x},\tau,t_0,t_1)+...\ . \label{sepdef}
\end{align}
Substitution into the non-local equation (\ref{non_loc}) gives the following at each order:
\begin{align}
O&(\epsilon^{-1})\qquad \qquad \frac{\partial \bm{u}_0}{\partial \tau}+\mathcal{L}(\bm{u}_0)=0, \label{lin_eqn} \qquad \qquad \qquad \qquad \ \ \ \\
O&(1)\qquad \qquad\ \ \ \frac{\partial \bm{u}_1}{\partial \tau}+\mathcal{L}(\bm{u}_1)=-\left(\frac{\partial \bm{u}_0}{\partial t_0}+\mathcal{N}(\bm{u}_0,\bm{u}_0)\right). \label{exp2} \\
O&(\epsilon)\qquad \qquad\ \ \  \frac{\partial \bm{u}_2}{\partial \tau}+\mathcal{L}(\bm{u}_2)=-\left(\frac{\partial \bm{u}_1}{\partial t_0}+\frac{\partial \bm{u}_0}{\partial t_1}+\mathcal{N}(\bm{u}_0,\bm{u}_1)+\mathcal{N}(\bm{u}_1,\bm{u}_0)\right). \label{exp3}
\end{align}
At first order (\ref{lin_eqn}) the equation is linear and we can write the solutions as:
\begin{eqnarray} \nonumber
\bm{u}_0(\bm{x},\tau,t_0,t_1)=e^{-\tau\mathcal{L}}\bar{\bm{u}}(\bm{x},t_0,t_1), \label{lin_sol}
\end{eqnarray}
or in terms of the eigenfunctions as given above in (\ref{four_bas}).
Solving at the next order (\ref{exp2}):
\begin{eqnarray}
\bm{u}_1e^{\tau\mathcal{L}}=\left. \bm{u}_1\right|_{\tau=0}-\left(\tau\frac{\partial \bar{\bm{u}}}{\partial t_0}+\int_0^\tau e^{s\mathcal{L}}\mathcal{N}(e^{-s\mathcal{L}}\bar{\bm{u}},e^{-s\mathcal{L}}\bar{\bm{u}})ds \right).
\end{eqnarray}
With the equation in this form we can identify possible secular terms as any of $O(\tau)$ or higher: those in the round brackets. To maintain the separation of scales for the velocities/pressures as defined in (\ref{sepdef}) these terms must be zero in the limit $\tau \rightarrow \infty$. This is the \lq{}cancellation of oscillations\rq{} concept, the generalised proof of convergence for general hyperbolic equations was given by \cite{schochet1994fast}. With the vector $\bar{\bm{u}}$ written in its eigenbasis such that the matrix exponential is just the exponential of the frequency of the corresponding eigenvalue ($e^{-i\omega t}$):
 \begin{align} \nonumber
\frac{\partial \bar{\bm{u}}}{\partial t_0}&=-\frac{1}{\tau}\int_0^\tau e^{s\mathcal{L}}\mathcal{N}(e^{-s\mathcal{L}}\bar{\bm{u}},e^{-s\mathcal{L}}\bar{\bm{u}})ds\\
=&-\frac{1}{\tau}\int_0^\tau \sum_{\substack{ \bm{k},\bm{k}_1,\bm{k}_2\\ \alpha,\alpha_1,\alpha_2\\ \bm{k}=\bm{k}_1+\bm{k}_2}}C_{\bm{k}_1\bm{k}_2\bm{k}}^{\alpha_1 \alpha_2 \alpha}a^{\alpha_1}_{\bm{k}_1}a^{\alpha_2 }_{\bm{k}_2}\bm{r}_{\bm{k}}^{\alpha  }e^{i\bm{k}\cdot\bm{x}}e^{i(\omega^{\alpha_1 }_{\bm{k}_1}+\omega^{\alpha_2  }_{\bm{k}_2}-\omega^{\alpha  }_{\bm{k}})s}ds, \label{rm_sec_terms}
 \end{align}
 where $a^{\alpha_i  }_{\bm{k}_i}$ represents the wave amplitude of each eigenfunction and the interaction coefficient is defined as:
 \begin{align} 
C_{\bm{k}_1\bm{k}_2\bm{k}}^{\alpha_1 \alpha_2 \alpha}=\left<\mathcal{N}(\bm{r}_{\bm{k}_1}^{\alpha_1},\bm{r}_{\bm{k}_2}^{\alpha_2}), \bm{r}_{\bm{k}}^{\alpha}\right> \label{1lintcoeff}
 \end{align}
 where $\left<\cdot,\cdot\right>$ is the scalar product of the system and derivatives in the operator $\mathcal{N}$ are expressed in the spectral space.
 
In the limit $\tau\rightarrow \infty$, the integral of all oscillatory contributions exactly cancel to 0 and so the only contributions come from the non-oscillatory constant contributions where $\omega^{\alpha_1  }_{\bm{k}_1}+\omega^{\alpha_2  }_{\bm{k}_2}-\omega^{\alpha }_{\bm{k}}=0$: the resonant triads. This leaves the equations:
\begin{align}
\frac{\partial \bar{\bm{u}}}{\partial t_0}
=& \sum_{\substack{ \bm{k},\bm{k}_1,\bm{k}_2\\ \alpha,\alpha_1,\alpha_2}}C_{\bm{k}_1\bm{k}_2\bm{k}}^{\alpha_1 \alpha_2 \alpha}a^{\alpha_1  m_1}_{\bm{k}_1}(t_0,t_1)a^{\alpha_2  }_{\bm{k}_2}(t_0,t_1)\bm{r}_{\bm{k}}^{\alpha }e^{i\bm{k}\cdot\bm{x}}\delta_{\bm{k}-\bm{k}_1-\bm{k}_2}\delta_{\omega-\omega_1-\omega_2}\ ,
\label{before_average}
 \end{align}
 or, in terms of only wave amplitudes:
 \begin{align}
\frac{\partial}{\partial t_0}a_{\bm{k}}^{\alpha }
=& \sum_{\substack{ \bm{k},\bm{k}_1,\bm{k}_2\\ \alpha,\alpha_1,\alpha_2}}C_{\bm{k}_1\bm{k}_2\bm{k}}^{\alpha_1 \alpha_2 \alpha}a^{\alpha_1 }_{\bm{k}_1}(t_0,t_1)a^{\alpha_2  }_{\bm{k}_2}(t_0,t_1)\delta_{\bm{k}-\bm{k}_1-\bm{k}_2}\delta_{\omega-\omega_1-\omega_2}\ . \label{1stclos}
 \end{align}
 
 This is the full expansion to triad order of the exact resonances. This is commonly extended to include the near resonances (see \cite{grimshaw2007nonlinear} for instance)-  those such that $\omega^{\alpha_1  }_{\bm{k}_1}+\omega^{\alpha_2  }_{\bm{k}_2}-\omega^{\alpha }_{\bm{k}}\sim \epsilon$. This is justified by considering the scale of the integral term in (\ref{rm_sec_terms}):
 \begin{align} \nonumber
\left|\frac{1}{\tau}\int_0^\tau e^{i\Omega s}ds\right|&= \left|\frac{e^{i\Omega\tau}-1}{\tau\Omega}\right|\le\frac{2}{\tau|\Omega|}\sim O\left(\frac{\epsilon}{\Omega}\right),\\ 
\text{where:}\qquad \Omega&=\omega^{\alpha  }_{\bm{k}}-\omega^{\alpha_1 }_{\bm{k}_1}-\omega^{\alpha_2  }_{\bm{k}_2}. \label{nr_argu}
\end{align}
Taking this weaker resonance condition we extend the definition of the resonant terms to include those at near resonance, and therefore include them in the equations at the triad order.

To understand the significance of this change to the asymptotic expansion we must continue to higher order in the following section. We will find that some of the interactions from higher order in the exact resonant expansion are promoted to higher order when the near resonant approximation is taken.

 \section{Quartet order expansion}
\label{quart_ord}
  To continue to higher order and form quartets we need to calculate the $\bm{u}_1$ terms, as these will be passed back into the quadratic nonlinearity to form a `pseudo' cubic interaction. When the nonlinearity contains no terms above quadratic order this process forms the whole of the higher order expansion. We define the term `slaved modes' as those belonging to the $\bm{u}_1$ part of the expansion as their time evolution is entirely derived from the $\bm{u}_0$ terms.
 \subsection{Slaved modes}
  \label{sec_u1_def}
We return to (\ref{exp2}):
\begin{align}
\frac{\partial \bm{u}_1}{\partial \tau}+\mathcal{L}(\bm{u}_1)=-\left(\frac{\partial \bm{u}_0}{\partial t_0}+\mathcal{N}(\bm{u}_0,\bm{u}_0)\right). 
\end{align}
We remove the exact and near-resonant terms that we have just equated with the $t_0$ derivative:
\begin{align}
\frac{\partial \bm{u}_1}{\partial \tau}+\mathcal{L}(\bm{u}_1)=-\mathcal{N}^{nr}(\bm{u}_0,\bm{u}_0). \label{slav_mode_eqn}
\end{align}
Here we have written $\mathcal{N}^{nr}$ to indicate that only the non-resonant parts of the nonlinear term are included. We now solve for $\bm{u}_1$. The complementary function would find linear modes of the same form as for $\bm{u}_0$ and hence we can assume they are identically 0 as the solution in this form is accounted for in those first order terms (see \cite{ablowitz2011nonlinear}, for example). Now we simply need to find the particular integral for our RHS:
\begin{align}
\bm{u}_1=-e^{-\tau\mathcal{L}}\int_0^\tau e^{s\mathcal{L}} \mathcal{N}^{nr}(e^{-s\mathcal{L}}\bar{\bm{u}},e^{-s\mathcal{L}}\bar{\bm{u}})ds \label{solve_for_slaved}
\end{align}
or in the eigenmode basis:
\begin{align} \nonumber
\{\bm{u}_1\}^{\alpha_a}_{\bm{k}_a}&=- \sum_{\substack{\bm{k}_1,\bm{k}_2\\ \alpha_1, \alpha_2}}\left.C_{\bm{k}_1\bm{k}_2\bm{k_a}}^{\alpha_1 \alpha_2 \alpha_a}\right|_{nr}a^{\alpha_1}_{\bm{k}_1} a^{\alpha_2}_{\bm{k}_2}\bm{r}^{\alpha_a}_{\bm{k}_a}e^{i\bm{k}\cdot\bm{x}}e^{-\omega_a\tau}\int_0^\tau e^{i(-\omega_1 -\omega_2 +\omega_a )s}ds\ \delta_{\bm{k}_a-\bm{k}_1-\bm{k}_2}\\
&=\sum_{\substack{\bm{k}_1,\bm{k}_2\\ \alpha_1, \alpha_2}}\frac{\left. C_{\bm{k}_1\bm{k}_2\bm{k_a}}^{\alpha_1 \alpha_2 \alpha_a}\right|_{nr}a^{\alpha_1}_{\bm{k}_1} a^{\alpha_2}_{\bm{k}_2}}{i(\omega_a-\omega_1-\omega_2)}\bm{r}^{\alpha_a}_{\bm{k}_a}e^{i\bm{k}\cdot\bm{x}}e^{-i(\omega_1 +\omega_2 )\tau}\ \delta_{\bm{k}_a-\bm{k}_1-\bm{k}_2} \label{slav_mod}
\end{align}
Note that $\bm{u}_1$ does {\bf not} evolve according to the dispersion relation: by the definition of non-resonance $\omega_a\ne\omega_1+\omega_2$. These are the slaved modes: they evolve only through the change to the two underlying modes $a^{\alpha_1}_{\bm{k}_1}$, $a^{\alpha_2}_{\bm{k}_2}$.

\subsection{Quartets as combinations of slaved modes}
\label{normal_quartets}
Now that we have the form of $\bm{u}_1$ from section \ref{sec_u1_def} we can continue to the next order of expansion in (\ref{exp3}):
\begin{align}
\frac{\partial \bm{u}_2}{\partial \tau}+\mathcal{L}(\bm{u}_2)=-\left(\frac{\partial \bm{u}_1}{\partial t_0}+\frac{\partial \bm{u}_0}{\partial t_1}+\mathcal{N}(\bm{u}_0,\bm{u}_1)+\mathcal{N}(\bm{u}_1,\bm{u}_0)\right).
\end{align}
Following the same process of removing secular terms that we used to derive equation (\ref{before_average}) we have the equation:
\begin{align}
\frac{\partial \bar{\bm{u}}}{\partial t_1}=-\lim_{\tau\rightarrow\infty}\frac{1}{\tau}\int_0^{\tau}e^{s\mathcal{L}}\left(e^{-s\mathcal{L}\rq{}}\frac{\partial \bar{\bm{u}}_1}{\partial t_0}+\mathcal{N}(e^{-s\mathcal{L}}\bar{\bm{u}},e^{-s\mathcal{L}\rq{}}\bar{\bm{u}}_1)+\mathcal{N}(e^{-s\mathcal{L}\rq{}}\bar{\bm{u}}_1,e^{-s\mathcal{L}}\bar{\bm{u}})\right)ds, \label{3part_eqn}
\end{align}
where $\bm{u}_1=e^{-\tau\mathcal{L}\rq{}}\bar{\bm{u}}_1$, expressed as $\{\bar{\bm{u}}_1\}_{\bm{k}}^{\alpha}e^{-i(\omega_1+\omega_2)\tau}$ in the eigenmode basis for some input modes subscripted $1$ and $2$. We now consider each term individually. Taking the first term on the right hand side:
\begin{align} \nonumber
-\lim_{\tau\rightarrow\infty}\frac{1}{\tau}&\int_0^{\tau}\left\{e^{s(\mathcal{L}-\mathcal{L}\rq{})}\frac{\partial \bar{\bm{u}}_1}{\partial t_0}\right\}_{\bm{k}}^{\alpha}\ ds\\
&=-\sum_{\substack{\bm{k}_1,\bm{k}_2\\ \alpha_1, \alpha_2}}\frac{\left. C_{\bm{k}_1\bm{k}_2\bm{k}}^{\alpha_1 \alpha_2 \alpha}\right|_{nr}}{i(\omega-\omega_1-\omega_2)}\ \delta_{\bm{k}-\bm{k}_1-\bm{k}_2}\lim_{\tau\rightarrow\infty}\frac{1}{\tau}\int_0^{\tau}\frac{\partial}{\partial t_0}(a^{\alpha_1}_{\bm{k}_1} a^{\alpha_2}_{\bm{k}_2})\bm{r}^{\alpha}_{\bm{k}}e^{i(\omega-\omega_1 -\omega_2 )s}ds.
\end{align}
Now from (\ref{1stclos}) the derivative in the integrand can be expanded:
\begin{align} \nonumber
\frac{\partial}{\partial t_0}(a^{\alpha_1}_{\bm{k}_1} a^{\alpha_2}_{\bm{k}_2})=&a^{\alpha_1}_{\bm{k}_1}\frac{\partial}{\partial t_0}( a^{\alpha_2}_{\bm{k}_2})+a^{\alpha_2}_{\bm{k}_2}\frac{\partial}{\partial t_0}(a^{\alpha_1}_{\bm{k}_1} )\\
&=a^{\alpha_1}_{\bm{k}_1}\sum_{\substack{\bm{k}_a,\bm{k}_b\\ \alpha_2,\alpha_a,\alpha_b\\ \bm{k}_2=\bm{k}_a+\bm{k}_b}}C_{\bm{k}_a\bm{k}_b\bm{k}_2}^{\alpha_a \alpha_b \alpha_2}a^{\alpha_a }_{\bm{k}_a}a^{\alpha_b  }_{\bm{k}_b}\ \ +\ \ a^{\alpha_2}_{\bm{k}_2}\sum_{\substack{\bm{k}_a,\bm{k}_b\\ \alpha_a,\alpha_b\\ \bm{k}_1=\bm{k}_a+\bm{k}_b}}C_{\bm{k}_a\bm{k}_b\bm{k}_1}^{\alpha_a \alpha_b \alpha_1}a^{\alpha_a }_{\bm{k}_a}a^{\alpha_b  }_{\bm{k}_b}.
\end{align}
None of these terms have any $\tau$ dependence, and so the derivative can be moved outside the integral and the limit:
\begin{align} \nonumber
-\lim_{\tau\rightarrow\infty}\frac{1}{\tau}&\int_0^{\tau}\left\{e^{s(\mathcal{L}-\mathcal{L}\rq{})}\frac{\partial \bar{\bm{u}}_1}{\partial t_0}\right\}_{\bm{k}}^{\alpha}\ ds\\
&=-\sum_{\substack{\bm{k}_1,\bm{k}_2\\ \alpha_1, \alpha_2}}\frac{\left. C_{\bm{k}_1\bm{k}_2\bm{k}}^{\alpha_1 \alpha_2 \alpha}\right|_{nr}}{i(\omega-\omega_1-\omega_2)}\ \delta_{\bm{k}-\bm{k}_1-\bm{k}_2}\frac{\partial}{\partial t_0}(a^{\alpha_1}_{\bm{k}_1} a^{\alpha_2}_{\bm{k}_2})\bm{r}^{\alpha}_{\bm{k}}\lim_{\tau\rightarrow\infty}\frac{1}{\tau}\int_0^{\tau}e^{i(\omega-\omega_1 -\omega_2 )s}ds.
\end{align}
We now have, similarly to the treatment of equation (\ref{before_average}) at the triad order, that the limit:
\begin{align}
\lim_{\tau\rightarrow\infty}\frac{1}{\tau}\int_0^{\tau}e^{i(\omega-\omega_1 -\omega_2 )s}ds,
\end{align}
will send every term to 0 by cancellation of oscillations, unless $\omega-\omega_1 -\omega_2=0$. We know that no terms have this form as we required the triad to be non-resonant in (\ref{slav_mode_eqn}) and hence this term must be 0.

We now consider the second term of (\ref{3part_eqn}). Writing in the eigenmode basis we see that upon substitution of (\ref{slav_mod}) we form various quartet resonances from the terms within the integral:
\begin{align} \nonumber
&-\lim_{\tau\rightarrow\infty}\frac{1}{\tau}\int_0^{\tau}\left\{e^{s\mathcal{L}}\mathcal{N}(e^{-s\mathcal{L}}\bar{\bm{u}},e^{-s\mathcal{L}\rq{}}\bar{\bm{u}}_1)\right\}^{\alpha}_{\bm{k}}ds\\ \nonumber
&=-\lim_{\tau\rightarrow\infty}\frac{1}{\tau}\int_0^{\tau}\sum_{\substack{\bm{k}_a,\bm{k}_3\\ \alpha_a, \alpha_3}}C_{\bm{k}_a\bm{k}_3\bm{k}}^{\alpha_a \alpha_3 \alpha}\bm{r}^{\alpha}_{\bm{k}}a_{\bm{k}_3}^{\alpha_3}e^{-i\omega_3s}\delta_{\bm{k}-\bm{k}_a-\bm{k}_3}\sum_{\substack{\bm{k}_1,\bm{k}_2\\ \alpha_1, \alpha_2}}\frac{\left. C_{\bm{k}_1\bm{k}_2\bm{k_a}}^{\alpha_1 \alpha_2 \alpha_a}\right|_{nr}a^{\alpha_1}_{\bm{k}_1} a^{\alpha_2}_{\bm{k}_2}}{i(\omega_a-\omega_1-\omega_2)}e^{-i(\omega_1 +\omega_2 )s}\ \delta_{\bm{k}_a-\bm{k}_1-\bm{k}_2}e^{i\omega s}ds\\ \nonumber
&=-\sum_{\substack{\bm{k}_a,\bm{k}_3\\ \alpha_a, \alpha_3}}\sum_{\substack{\bm{k}_1,\bm{k}_2\\ \alpha_1, \alpha_2}}\lim_{\tau\rightarrow\infty}\frac{1}{\tau}\int_0^{\tau}\frac{C_{\bm{k}_a\bm{k}_3\bm{k}}^{\alpha_a \alpha_3 \alpha}\left. C_{\bm{k}_1\bm{k}_2\bm{k_a}}^{\alpha_1 \alpha_2 \alpha_a}\right|_{nr}}{i(\omega_a-\omega_1-\omega_2)}a^{\alpha_1}_{\bm{k}_1} a^{\alpha_2}_{\bm{k}_2}a_{\bm{k}_3}^{\alpha_3}\bm{r}^{\alpha}_{\bm{k}}e^{i(\omega-\omega_1 -\omega_2-\omega_3)s}\ \delta_{\bm{k}-\bm{k}_1-\bm{k}_2-\bm{k}_3}ds\\
&=-\sum_{\substack{\bm{k}_1,\bm{k}_2,\bm{k}_3\\ \alpha_a,\alpha_1, \alpha_2, \alpha_3}}\frac{C_{\bm{k}_a\bm{k}_3\bm{k}}^{\alpha_a \alpha_3 \alpha}\left. C_{\bm{k}_1\bm{k}_2\bm{k_a}}^{\alpha_1 \alpha_2 \alpha_a}\right|_{nr}}{i(\omega_a-\omega_1-\omega_2)}a^{\alpha_1}_{\bm{k}_1} a^{\alpha_2}_{\bm{k}_2}a_{\bm{k}_3}^{\alpha_3}\bm{r}^{\alpha}_{\bm{k}}\delta_{\omega-\omega_1 -\omega_2-\omega_3}\ \delta_{\bm{k}-\bm{k}_1-\bm{k}_2-\bm{k}_3}. \label{qtets}
\end{align}
By the symmetry of the nonlinear interaction coefficient the third term of (\ref{3part_eqn}) is identical to the second and so we can now write the whole equation as: 
\begin{align}
\frac{\partial\ }{\partial t_1}\{\bar{\bm{u}}_0\}_{\bm{k}}^{\alpha}&=-\sum_{\substack{\bm{k}_1,\bm{k}_2,\bm{k}_3\\ \alpha_1, \alpha_2, \alpha_3}}Q_{\bm{k}_1\bm{k}_2\bm{k}_3\bm{k}}^{ \alpha_1 \alpha_2 \alpha_3\alpha}a^{\alpha_1}_{\bm{k}_1} a^{\alpha_2}_{\bm{k}_2}a_{\bm{k}_3}^{\alpha_3}\bm{r}^{\alpha}_{\bm{k}}\delta_{\omega-\omega_1 -\omega_2-\omega_3}\ \delta_{\bm{k}-\bm{k}_1-\bm{k}_2-\bm{k}_3}, \label{qtet_eqn}
\end{align}
where:
\begin{align}
Q_{\bm{k}_1\bm{k}_2\bm{k}_3\bm{k}}^{ \alpha_1 \alpha_2 \alpha_3\alpha}=\sum_{\alpha_a}\frac{2C_{\bm{k}_a\bm{k}_3\bm{k}}^{\alpha_a \alpha_3 \alpha}\left. C_{\bm{k}_1\bm{k}_2\bm{k_a}}^{\alpha_1 \alpha_2 \alpha_a}\right|_{nr}}{i(\omega_a-\omega_1-\omega_2)}, \label{sing_quart}
\end{align}
and $\bm{k_a}=\bm{k_1}+\bm{k_2}$ and $\bm{k_a}=\bm{k}-\bm{k_3}$.

Similarly to the triad case it is useful to symmetrise this in the input coefficients (1,2,3) as follows:
\begin{align} 
Q_{\bm{k}_1\bm{k}_2\bm{k}_3\bm{k}}^{ \alpha_1 \alpha_2 \alpha_3\alpha}=\frac{2}{3}\left[ \sum_{\alpha_a}\frac{C_{\bm{k}_a\bm{k}_3\bm{k}}^{\alpha_a \alpha_3 \alpha}\left. C_{\bm{k}_1\bm{k}_2\bm{k_a}}^{\alpha_1 \alpha_2 \alpha_a}\right|_{nr}}{i(\omega_a-\omega_1-\omega_2)}
+\sum_{\alpha_b}\right.&\frac{C_{\bm{k}_b\bm{k}_1\bm{k}}^{\alpha_b \alpha_1 \alpha}\left. C_{\bm{k}_2\bm{k}_3\bm{k_b}}^{\alpha_2 \alpha_3 \alpha_b}\right|_{nr}}{i(\omega_b-\omega_2-\omega_3)}  \left.
+\sum_{\alpha_c}\frac{C_{\bm{k}_c\bm{k}_2\bm{k}}^{\alpha_c \alpha_2 \alpha}\left. C_{\bm{k}_3\bm{k}_1\bm{k_c}}^{\alpha_3 \alpha_1 \alpha_c}\right|_{nr}}{i(\omega_c-\omega_3-\omega_1)}\right]. \label{Qdef}
\end{align}

\begin{figure}[htb]
\centerline{\includegraphics[width=8cm]{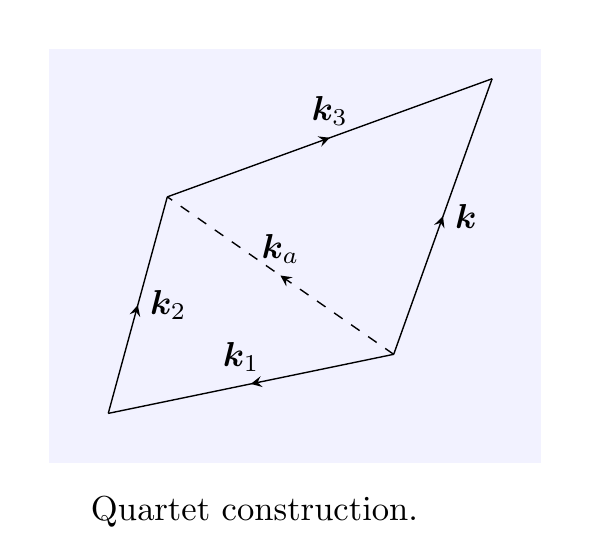}}
\caption{Quartet construction, as in Equation (\ref{sing_quart}). The wave vector $\bm{k}_a$ takes the form of a slaved mode that can be projected onto our eigenbasis such that there are fast-like parts and a slow-like part. Each quartet has three possible sets of wavenumbers that can contribute to the quartet, ie $\bm{k}_a=\bm{k}_1+\bm{k}_2$, $\bm{k}_b=\bm{k}_3+\bm{k}_2$, and $\bm{k}_c=\bm{k}_1+\bm{k}_3$.}
\label{quart_diag}
\end{figure}

Figure \ref{quart_diag} shows diagrammatically how the non-resonant triads combine to form a quartet.\\
\\

This completes the asymptotic expansion up to the quartet order of expansion, defining all dynamics up to the $t_1$ timescale. It should be noted that this analysis would appear almost identical if we had taken the exact or near resonant triad condition at the previous order, the difference is simply in which terms are present in $\mathcal{N}^{nr}$ in equation \ref{slav_mode_eqn} and elsewhere. However although this appears as a small change in the written analysis the effect on the behaviour is substantial. We can show that the terms that result from near resonant triads are in fact exactly equivalent to the slaved modes, and that combinations of them form quartets in the same manner, but now promoted to the same timescale as the triad interactions.

\subsection{Fast quartets as combinations of near resonances}
\label{nrsec_sec_nr_exp}
Literature describing isolated near resonant triads can be found starting originally in \cite{bretherton1964resonant} and more recently in \cite{grimshaw2007nonlinear}. In the isolated case not much is changed from exact resonant interactions, the detuning can only affect the phase by an $O(1)$ amount and changes to the behaviour of the amplitude can only be $O(\epsilon)$. Here we consider the triads in a non-isolated manner, such that they may interact with one another as permitted by the triad condition $\bm{k}=\bm{k}_1+\bm{k}_2$. This will lead to qualitatively different behaviour compared to the exact resonant expansion.

To see this, we first return to (\ref{solve_for_slaved}), now written in the eigenmode basis:
 \begin{align} \nonumber
\frac{\partial \bar{\bm{u}}}{\partial t_0}=&-\frac{1}{\tau}\int_0^\tau \sum_{\substack{ \bm{k},\bm{k}_1,\bm{k}_2\\ \alpha,\alpha_1,\alpha_2\\ \bm{k}=\bm{k}_1+\bm{k}_2}}C_{\bm{k}_1\bm{k}_2\bm{k}}^{\alpha_1 \alpha_2 \alpha}a^{\alpha_1}_{\bm{k}_1}(t_0,t_1)a^{\alpha_2 }_{\bm{k}_2}(t_0,t_1)\bm{r}_{\bm{k}}^{\alpha  }e^{i\bm{k}\cdot\bm{x}}e^{i\Omega_{\bm{k}_1\bm{k}_2\bm{k}}^{\alpha_1 \alpha_2 \alpha}s}ds.
 \end{align}
 However we now explicitly write the $\epsilon$ scaling of $\Omega$ such that $\tau\Omega=\tau\epsilon\Delta=t_0\Delta$, with $\Delta=O(1)$ which slows the dependence of the exponential to the $t_0$ timescale:
 \begin{align} \nonumber
\frac{\partial a_{\bm{k}}^{\alpha}}{\partial t_0}=&- \sum_{\substack{ \bm{k}_1,\bm{k}_2\\ \alpha_1,\alpha_2\\ \bm{k}=\bm{k}_1+\bm{k}_2}}C_{\bm{k}_1\bm{k}_2\bm{k}}^{\alpha_1 \alpha_2 \alpha}a^{\alpha_1}_{\bm{k}_1}(t_0,t_1)a^{\alpha_2 }_{\bm{k}_2}(t_0,t_1)e^{i\Delta_{\bm{k}_1\bm{k}_2\bm{k}}^{\alpha_1 \alpha_2 \alpha}t_0}\frac{1}{\tau}\int_0^\tau ds\\
=&- \sum_{\substack{ \bm{k}_1,\bm{k}_2\\ \alpha_1,\alpha_2\\ \bm{k}=\bm{k}_1+\bm{k}_2}}C_{\bm{k}_1\bm{k}_2\bm{k}}^{\alpha_1 \alpha_2 \alpha}a^{\alpha_1}_{\bm{k}_1}(t_0,t_1)a^{\alpha_2 }_{\bm{k}_2}(t_0,t_1)e^{i\Delta_{\bm{k}_1\bm{k}_2\bm{k}}^{\alpha_1 \alpha_2 \alpha}t_0}. \label{nrsec_nr_diff_eqn}
 \end{align}
 This dynamical equation is similar to that of the exact resonances (\ref{1stclos}), except for the exponential term. It defines evolution on the $t_0$ timescale, as before. 
 
For use later we now also perform the integration in $t_0$. This is similar to solving for the slaved modes, except on the slower timescale $t_0$ (and larger $\bm{u}_0$). This is done by parts:
  \begin{align} \nonumber
a_{\bm{k}}^{\alpha}
=- \sum_{\substack{ \bm{k}_1,\bm{k}_2\\ \alpha_1,\alpha_2\\ \bm{k}=\bm{k}_1+\bm{k}_2}}&C_{\bm{k}_1\bm{k}_2\bm{k}}^{\alpha_1 \alpha_2 \alpha}\left(\frac{a^{\alpha_1}_{\bm{k}_1}(t_0,t_1)a^{\alpha_2 }_{\bm{k}_2}(t_0,t_1)}{i\Delta_{\bm{k}_1\bm{k}_2\bm{k}}^{\alpha_1 \alpha_2 \alpha}}e^{i\Delta_{\bm{k}_1\bm{k}_2\bm{k}}^{\alpha_1 \alpha_2 \alpha}t_0}\right.\\
&-\left.\int_0^{t_0}\frac{\frac{\partial }{\partial s}(a^{\alpha_1}_{\bm{k}_1}(s,t_1))a^{\alpha_2 }_{\bm{k}_2}(s,t_1)+a^{\alpha_1}_{\bm{k}_1}(s,t_1)\frac{\partial }{\partial s}(a^{\alpha_2 }_{\bm{k}_2}(s,t_1))}{i\Delta_{\bm{k}_1\bm{k}_2\bm{k}}^{\alpha_1 \alpha_2 \alpha}}e^{i\Delta_{\bm{k}_1\bm{k}_2\bm{k}}^{\alpha_1 \alpha_2 \alpha}s}ds\right). \label{nrsec_nr_first_exp}
 \end{align}
 We will pause the analysis to mention two things. Firstly, the left hand term in the sum has the same form as the slaved modes, except for the timescale of the exponential. Secondly the right hand term can be integrated again in the same manner, which we will see mirrors the quartet expansion process. For this reason the $t_1$ dependence has been explicitly written in the arguments of the amplitudes.
 
 The key observation is that the near resonances behave exactly as the slaved mode interactions, except that instead of affecting the $\bm{u}_1$ part on timescale $\tau$ they affect the $\bm{u}_0$ part on timescale $t_0$. This allows them to evolve independently; they are not behaving in the slaved manner of the $\bm{u}_1$ part.
 
To form the next `order' of the near resonant expansion we will follow what is effectively the same process again, although there are now key differences in the mathematics, notably in the timescales. We assume that at least one of the amplitudes in the RHS of (\ref{nrsec_nr_diff_eqn}) is also a member of another near resonant triad and substitute (\ref{nrsec_nr_first_exp}) in its place:
  \begin{align} \nonumber
\frac{\partial a_{\bm{k}}^{\alpha}}{\partial t_0}=&- \sum_{\substack{\bm{k}_3, \bm{k}_a \\ \alpha_3, \alpha_a\\ \bm{k}=\bm{k}_3+\bm{k}_a}}\sum_{\substack{ \bm{k}_1,\bm{k}_2\\ \alpha_1,\alpha_2\\ \bm{k}_a=\bm{k}_1+\bm{k}_2}}\frac{C_{\bm{k}_3\bm{k}_a\bm{k}}^{\alpha_3 \alpha_a \alpha}C_{\bm{k}_1\bm{k}_2\bm{k}_a}^{\alpha_1 \alpha_2 \alpha_a}}{i\Delta_{\bm{k}_1\bm{k}_2\bm{k}_a}^{\alpha_1 \alpha_2 \alpha_a}}a^{\alpha_3}_{\bm{k}_3}a^{\alpha_1}_{\bm{k}_1}a^{\alpha_2 }_{\bm{k}_2}e^{i\Delta_{\bm{k}_1\bm{k}_2\bm{k}_a}^{\alpha_1 \alpha_2 \alpha_a}t_0}e^{i\Delta_{\bm{k}_3\bm{k}_a\bm{k}}^{\alpha_3 \alpha_a \alpha}t_0}\\ \label{nrsec_nr_quarts}
& \qquad \qquad \qquad \qquad \qquad \qquad \qquad+ \text{above quartet terms} ,
 \end{align}
 where the right hand term of (\ref{nrsec_nr_first_exp}) gets pushed into the \lq{}above quartet terms\rq{} because the $\partial/\partial s$ terms will be functions of two or more amplitudes as shown by (\ref{nrsec_nr_diff_eqn}). 
 
 This again mirrors the exact resonant quartet expansion (the same interaction coefficient Q as given in (\ref{sing_quart}), would be formed, although now scaled by $\epsilon$), except that the evolution is still on the $t_0$ timescale. Because of our imposed structure we should note that the near resonant triads can only form a \lq{}higher order interaction\rq{} with other near resonant sets, and slaved modes can only form a higher order resonance by combination with each other.
 
The near resonant expansion (that evolves according to dynamical equation (\ref{nrsec_nr_diff_eqn})) contains a subset of the exact quartet resonances from the analysis of section \ref{normal_quartets}, which behave as before, but now acting on the faster $t_0$ timescale instead of $t_1$. The near resonant expansion contains the strongest of the exact quartet resonances, constructed from near resonant triads, without proceeding to a higher order.

One way to interpret this is to say that in forming the exact quartet interaction coefficient given by equation (\ref{Qdef}) we divide by $\Omega$, but in the case of the near resonant component triads this causes an extremely large interaction coefficient $\sim1/\epsilon$. The near resonant interactions pick out the cases where the higher order interaction coefficient is large enough to rearrange the asymptotic ordering.

\section{Expansion to higher order}
\label{high_ord}
In a similar manner of linking triads together one can form \lq{}n-tets\rq{}, a term we introduce as the generalisation of triads, quartets, quintets, sextets etc, for combinations of $n$ modes. In the following we establish a general form for the mathematics of these interactions, from which we can deduce results that hold to all orders of expansion. We start by assuming an exact expansion, then show how once again the near resonant version of the expansion constitutes a reshuffling of the faster acting terms to the triad order.

\subsection{N-tets as a combination of slaved modes}
\label{ntet_exp}
The n-tets are defined by the pth order equation:
\begin{align}
O&(\epsilon^{p})\qquad \qquad\ \ \  \frac{\partial \bm{u}_{p+1}}{\partial \tau}+\mathcal{L}\bm{u}_{p+1}=-\sum_{m=0}^{p}\left(\frac{\partial \bm{u}_{p-m}}{\partial t_m}+\mathcal{N}(\bm{u}_m,\bm{u}_{p-m})\right), \label{gen_exp}
\end{align}
the general form of equations (\ref{lin_eqn})-(\ref{exp3}), where $p=n-3$.

\begin{table}[h]
\centerline{
\begin{tabular}{c|c|c|c|c|c|c}
\qquad\qquad\qquad& $\bm{u}_0$ & $\bm{u}_1$ & $\bm{u}_2$ & $\bm{u}_3$ & $\bm{u}_4$ & ...\\
 \hline 
$O(1/\epsilon)$ & $\tau$ & - & - & - & - & ...\\
$O(1)$ & $t_0$ & $\tau$ & - & - & - & ...\\
$O(\epsilon)$ & $t_1$ & $\color{red}t_0$ & $\tau$  & - & -& ...\\
$O(\epsilon^2)$ & $t_2$ &\color{red} $t_1$ &\color{red} $t_0$& $\tau$ &-& ...\\
$O(\epsilon^3)$ & $t_3$ &\color{red} $t_2$ &\color{red} $t_1$ &\color{red} $t_0$& $\tau$ & ...\\
\vdots &\vdots &\vdots &\vdots &\vdots &\vdots &  $\ddots$
\end{tabular}}
\caption{For given order of the equations, the timescale to which each order of the solution is considered is given. The timescales given in red are completely determined as they are simply derived from slaved modes and so their behaviour is inherited from the behaviour of $\bm{u}_0$ to the required timescale.}
\label{ntet_tab}
\end{table}

\begin{figure}[htb]
\centerline{\includegraphics[width=16cm]{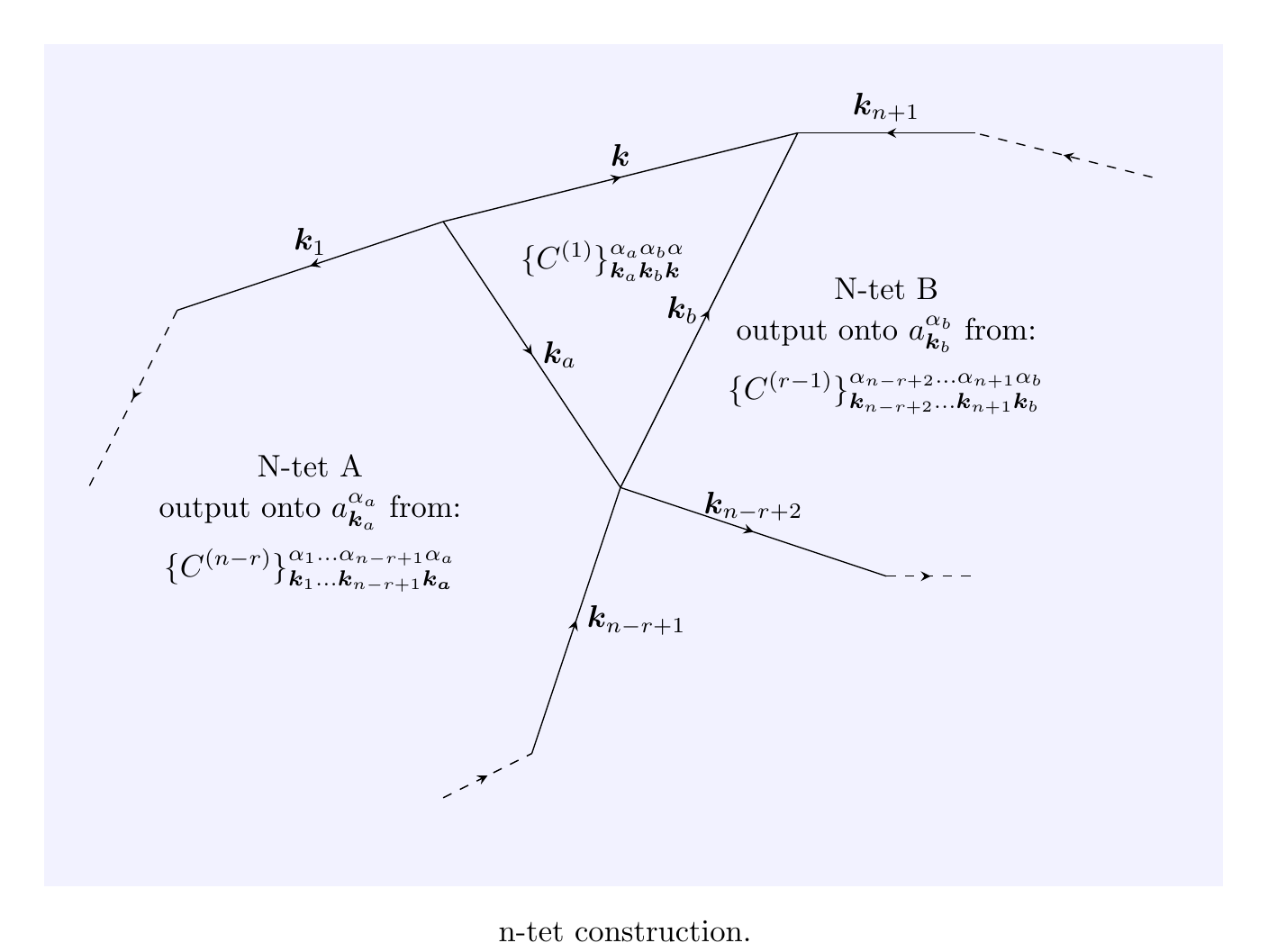}}
\caption{Diagram of the n-tet construction, Two smaller N-tets, A\&B of sizes $n-r+2$ and $r+1$ respectively, output the two modes $a$, and $b$ that go into a connecting triad that returns the output mode with wavenumber $\bm{k}$. All possible combinations must be considered in a similar manner, summed, and then the coefficient symmetrised to give the equation in (\ref{nth_ord_coeff}) }
\label{ntet_diag}
\end{figure}

We consider table \ref{ntet_tab}, which breaks down the dependence of each order of the asymptotic series in terms of its dependence on the various timescales. The terms in red are those which are already determined by the behaviour of $\bm{u}_0$, ie slaved. Hence, in the same manner with which we justified that the left hand term in (\ref{3part_eqn}) was negligible, our secularity condition will never contain any of the time derivatives in (\ref{gen_exp}) other than that of ${\partial \bm{u}_0}/{\partial t_p}$. Crucially, this means that the terms defining the evolution on the $t_{p}$ timescale will always reduce to just the nonlinear combinations $\mathcal{N}(\bm{u}_a,\bm{u}_b)$ for all non-negative integers $a,b$ such that $p=a+b$. The dynamics of each of the contributing components $\bm{u}_a,\bm{u}_b$ can be calculated as a function of those at lower orders which in turn behave in the same way, until all motion is determined by the $\bm{u}_0$ terms. We can therefore calculate a general higher order interaction coefficient as a recurrence relation. \\
\\
We start by calculating the different possible combinations $a$, $b$ available for the nonlinear part $\mathcal{N}(\bm{u}_a,\bm{u}_b)$. Given some input $\bm{u}_a$ this must be formed of $a+1$ $\bm{u}_0$ modes in an $a+2$-tet, and similarly for $\bm{u}_b$. In the output triad we have $a+b+2$ input $\bm{u}_0$ modes that must be split into the two groups of sizes $a+1,\ b+1$. Define $n=a+b+1$ so that n is the order of interaction (n=1 corresponds to triads), and $r=b+1$, the number of input modes into the second triad, the first triad then has $n-r+1=a+1$ input modes. The number of possible mode permutations then corresponds to $(n+1)!$ where these are shared between the two input n-tets in the division determined by the value of $r$.\\
\\
To symmetrise the interaction coefficient we must add all permutations of the input wavenumbers and then divide by the number of them, $(n+1)!$. Due to symmetries in the input interaction coefficients there will be repeated contributions ie $C_{\bm{k}_1\bm{k}_2\bm{k_a}}^{\alpha_1 \alpha_2 \alpha_a}$ and $C_{\bm{k}_2\bm{k}_1\bm{k_a}}^{\alpha_2 \alpha_1 \alpha_a} $ will both appear in different terms of the relation, although both are equal. However the relation is most simply stated in the form given below.\\
\\
For each of the permutations the quantity required will be the multiplication of the two input n-tet coefficients with a triad coefficient that takes these two outputs as its inputs and returns the output. This manner of combining smaller n-tets is displayed diagrammatically in figure \ref{ntet_diag}. We then need to sum over each possible $r$, to give all possible splittings into two n-tets.\\
\\
Combining these components we get the interaction coefficients, to any desired order, defined inductively as follows:
\begin{align} \nonumber
\{C&^{(n)}\}_{\bm{k}_1...\bm{k}_{n+1}\bm{k}}^{\alpha_1... \alpha_{n+2} \alpha}=\\ \nonumber
&\frac{1}{(n+1)!}\sum_{r=1}^{n}\Bigg[ \sum_{\alpha_a,\alpha_b}\frac{ \{C^{(1)}\}_{\bm{k}_a\bm{k}_b\bm{k}}^{\alpha_a \alpha_b \alpha}\left. \{C^{(r-1)}\}_{\bm{k}_{n-r+2}...\bm{k}_{n+1}\bm{k}_b}^{\alpha_{n-r+2}...\alpha_{n+1} \alpha_b}\right|_{nr}\left. \{C^{(n-r)}\}_{\bm{k}_1...\bm{k}_{n-r+1}\bm{k_a}}^{\alpha_1... \alpha_{n-r+1} \alpha_a}\right|_{nr}}{-\{\Omega^{(n-r)}\}^{a}_{1,...,n-r+1}\{\Omega^{(r-1)}\}^{b}_{n-r+2,...,n+1}}\\ &\qquad\qquad\qquad\qquad\qquad\qquad\qquad\qquad\qquad\qquad\qquad\qquad + \textit{input wavenumber permutations}\Bigg],\label{nth_ord_coeff}\\
&\text{where:}\qquad\{\Omega^{(n)}\}^{a}_{b,...,c}=\omega_a-\omega_{b}-...-\omega_{c}. \nonumber
\end{align}
Where $C^{(n)}$ is the nonlinear interaction coefficient at the nth closure (ie n=1 would be triad interactions at the first closure). We need to define the particular case of $\{C^{(0)}\}_{\bm{k}_i\bm{k}}^{\alpha_i \alpha}=1$ such that it is simply an identity mapping with $\bm{k}_i=\bm{k}$, $\alpha_i= \alpha$. We also define separately $\Omega^{(0)}=-i$. These special cases are necessary terms corresponding to $\mathcal{N}(\bm{u}_0,\bm{u}_{n-1})$ where an n-1-tet is combined with a mode, not another n-tet. This allows one to algorithmically compute a given interaction coefficient to any order.

\subsection{Fast n-tets as a combination of near resonances}
 As occurred for quartets in section \ref{nrsec_sec_nr_exp}, the near resonances will expand in the same manner as the exact resonant part, except that the n-tets, for every order n are all evolving on the $t_0$ timescale. To show the equivalent higher order behaviour in the near resonances we will continue the expansion using integration by parts, as we did in section \ref{nrsec_sec_nr_exp}.\\
\\
We take one of the right hand terms (with minor relabelling) of (\ref{nrsec_nr_first_exp}) and substitute into (\ref{nrsec_nr_diff_eqn}) to replace the derivative in $t_0$:
\begin{align} \nonumber
&\sum_{\substack{ \bm{k}_1,\bm{k}_a\\ \alpha_1,\alpha_a\\ \bm{k}=\bm{k}_1+\bm{k}_a}}C_{\bm{k}_1\bm{k}_a\bm{k}}^{\alpha_1 \alpha_a \alpha}\int_0^{t_0}\frac{\frac{\partial }{\partial s}(a^{\alpha_a}_{\bm{k}_a}(s,t_1))a^{\alpha_1 }_{\bm{k}_1}(s,t_1)}{i\Delta_{\bm{k}_1\bm{k}_a\bm{k}}^{\alpha_1 \alpha_a \alpha}}e^{i\Delta_{\bm{k}_1\bm{k}_a\bm{k}}^{\alpha_1 \alpha_a \alpha}s}ds\\
&=\sum_{\substack{\bm{k}_1,\bm{k}_a\\ \alpha_1,\alpha_a\\ \bm{k}=\bm{k}_1+\bm{k}_a}}\sum_{\substack{ \bm{k}_2,\bm{k}_3\\ \alpha_2,\alpha_3\\ \bm{k}_a=\bm{k}_2+\bm{k}_3}}\frac{C_{\bm{k}_1\bm{k}_a\bm{k}}^{\alpha_1 \alpha_a \alpha}C_{\bm{k}_2\bm{k}_3\bm{k}_a}^{\alpha_2 \alpha_3 \alpha_a}}{i\Delta_{\bm{k}_1\bm{k}_a\bm{k}}^{\alpha_1 \alpha_a \alpha}}\int_0^{t_0}a^{\alpha_1}_{\bm{k}_1}(s,t_1)a^{\alpha_2}_{\bm{k}_2}(s,t_1)a^{\alpha_3}_{\bm{k}_3}(s,t_1)e^{i\Delta_{\bm{k}_1\bm{k}_a\bm{k}}^{\alpha_1 \alpha_a \alpha}s}e^{i\Delta_{\bm{k}_2\bm{k}_3\bm{k}_a}^{\alpha_2 \alpha_3 \alpha_a}s}ds. 
 \end{align}
 This process would then be continued by repeatedly expanding the integral by parts, each time the integral remainder treated similarly, forming an expansion containing all combinations of n amplitudes (for $n>2$) that obey the n-tet condition $\sum_{i=1}^{n-1}\bm{k}_i=\bm{k}$. This expansion is then substituted back into the equation (\ref{nrsec_nr_first_exp}) in every place that an amplitude is involved in another near resonance. This will return, after some more relabelling, (\ref{nrsec_nr_quarts}) except that we can now see the form of the previously omitted \lq{}above quartet terms\rq{}:
   \begin{align} \nonumber
\frac{\partial a_{\bm{k}}^{\alpha}}{\partial t_0}=&- \sum_{\substack{\bm{k}_3, \bm{k}_a \\ \alpha_3, \alpha_a\\ \bm{k}=\bm{k}_3+\bm{k}_a}}\sum_{\substack{ \bm{k}_1,\bm{k}_2\\ \alpha_1,\alpha_2\\ \bm{k}_a=\bm{k}_1+\bm{k}_2}}\frac{C_{\bm{k}_3\bm{k}_a\bm{k}}^{\alpha_3 \alpha_a \alpha}C_{\bm{k}_1\bm{k}_2\bm{k}_a}^{\alpha_1 \alpha_2 \alpha_a}}{i\Delta_{\bm{k}_1\bm{k}_2\bm{k}_a}^{\alpha_1 \alpha_2 \alpha_a}}a^{\alpha_3}_{\bm{k}_3}a^{\alpha_1}_{\bm{k}_1}a^{\alpha_2 }_{\bm{k}_2}e^{i\Delta_{\bm{k}_1\bm{k}_2\bm{k}_a}^{\alpha_1 \alpha_2 \alpha_a}t_0}e^{i\Delta_{\bm{k}_3\bm{k}_a\bm{k}}^{\alpha_3 \alpha_a \alpha}t_0}\\ \nonumber
&- \sum_{\substack{\bm{k}_b, \bm{k}_a \\ \alpha_b, \alpha_a\\ \bm{k}=\bm{k}_b+\bm{k}_a}}\sum_{\substack{ \bm{k}_1,\bm{k}_2\\ \alpha_1,\alpha_2\\ \bm{k}_a=\bm{k}_1+\bm{k}_2}}\sum_{\substack{ \bm{k}_3,\bm{k}_4\\ \alpha_3,\alpha_4\\ \bm{k}_b=\bm{k}_3+\bm{k}_4}}\frac{C_{\bm{k}_b\bm{k}_a\bm{k}}^{\alpha_b \alpha_a \alpha}C_{\bm{k}_1\bm{k}_2\bm{k}_a}^{\alpha_1 \alpha_2 \alpha_a}C_{\bm{k}_3\bm{k}_4\bm{k}_b}^{\alpha_3 \alpha_4 \alpha_b}}{-\Delta_{\bm{k}_3\bm{k}_4\bm{k}_b}^{\alpha_3 \alpha_4 \alpha_b}\Delta_{\bm{k}_1\bm{k}_2\bm{k}_a}^{\alpha_1 \alpha_2 \alpha_a}}a^{\alpha_1}_{\bm{k}_1}a^{\alpha_2 }_{\bm{k}_2}a^{\alpha_3}_{\bm{k}_3}a^{\alpha_4}_{\bm{k}_4}e^{i\Delta_{\bm{k}_1\bm{k}_2\bm{k}_3\bm{k}_4\bm{k}}^{\alpha_1 \alpha_2 \alpha_3 \alpha_4 \alpha}t_0}\\
&- \sum_{\substack{\bm{k}_4, \bm{k}_a \\ \alpha_4, \alpha_a\\ \bm{k}=\bm{k}_4+\bm{k}_a}}\sum_{\substack{ \bm{k}_1,\bm{k}_2\bm{k}_3\\ \alpha_1,\alpha_2\alpha_3\\ \bm{k}_a=\bm{k}_1+\bm{k}_2}+\bm{k}_3}\frac{C_{\bm{k}_4\bm{k}_a\bm{k}}^{\alpha_4 \alpha_a \alpha}\{C^{(2)}\}_{\bm{k}_1\bm{k}_2\bm{k}_3\bm{k}_a}^{\alpha_1 \alpha_2 \alpha_3 \alpha_a}}{i\Delta_{\bm{k}_1\bm{k}_2\bm{k}_3\bm{k}_a}^{\alpha_1 \alpha_2 \alpha_3 \alpha_a}}a^{\alpha_1}_{\bm{k}_1}a^{\alpha_2 }_{\bm{k}_2}a^{\alpha_3}_{\bm{k}_3}a^{\alpha_4}_{\bm{k}_4}e^{i\Delta_{\bm{k}_1\bm{k}_2\bm{k}_3\bm{k}_4\bm{k}}^{\alpha_1 \alpha_2 \alpha_3 \alpha_4 \alpha}t_0}+...\ .\label{nrsec_above_quart}
 \end{align}
This is exactly as for the general higher order coefficient (\ref{nth_ord_coeff}), except for some $\epsilon$ scaling. If one were to continue this, always substituting in for near resonant terms, then the expansion will eventually contain only exact and super-near resonances. All of these will look identical in form to a subset of the exact resonant expansion, but will take place on the $t_0$ timescale. \\
 \\
This expansion is not necessary in the practical use of the near resonant expansion. It is simply a method to show that the near resonant triads interact exactly like the strongest parts of a full exact expansion. All of the dynamics is captured simply by maintaining the near resonances in the triad equations given in (\ref{nrsec_nr_diff_eqn}).

\subsection{Splitting of n-tets in the near resonant part}
\label{nrsec_ntet_split}
 Interestingly, n-tets can be split between orders in the near resonant expansion. As an example, consider the rotating shallow water equations with small Coriolis parameter $f$ (where fast modes have $\omega\sim \alpha c|\bm{k}|$) for the following quartet of modes:
 \begin{subequations}
\begin{align}
&\bm{k}_1, &\alpha_1=+,\\  
&\bm{k}_2=\bm{k}_1^{\perp}, &\alpha_2=+,\\ 
&\bm{k}_3=-\bm{k}_1+\epsilon^2\bm{k}_1^{\perp}, &\alpha_3=-,\\
&\bm{k}=\bm{k}_1^{\perp}(1+\epsilon^2), &\alpha=+.
\end{align}
\end{subequations}
This forms a near resonant quartet. From these we calculate the slaved mode wavenumbers that would contribute to the quartet:
 \begin{subequations}
\begin{align}
&\bm{k}_a=\bm{k}_1+\bm{k}_2=\bm{k}_1+\bm{k}_1^\perp,\\  
&\bm{k}_b=\bm{k}_2+\bm{k}_3=-\bm{k}_1+(1+\epsilon^2)\bm{k}_1^{\perp},\\ 
&\bm{k}_c=\bm{k}_3+\bm{k}_1=\epsilon^2\bm{k}_1^{\perp}.
\end{align}
\end{subequations}
 Calculating $\Omega$ in each case (assuming all fast modes):
\begin{subequations}
\begin{align}
&\Omega_{12a}=\omega(\bm{k}_1+\bm{k}_1^\perp)-\omega(\bm{k}_1)-\omega(\bm{k}_1^{\perp})\approx c(\sqrt{2}-2)|\bm{k}_1|,\\  
&\Omega_{23b}=\omega(-\bm{k}_1+(1+\epsilon^2)\bm{k}_1^{\perp})-\omega(\bm{k}_1)+\omega(-\bm{k}_1+\epsilon^2\bm{k}_1^{\perp})\approx c(\sqrt{2})|\bm{k}_1|,\\ 
&\Omega_{13c}=\omega(\epsilon^2\bm{k}_1^{\perp})+\omega(\bm{k}_1)-\omega(-\bm{k}_1+\epsilon^2\bm{k}_1^{\perp})\approx c\epsilon^2|\bm{k}_1|.
\end{align}
\end{subequations}
 
We see that a fast mode $c$ will be part of a near resonant triad. However fast-like mode $a$ or $b$ would not be included in the near resonant triads.

\section{Implications for zero mode systems}
\label{zero_mode_implic}
Having put the interaction coefficient in a general form, we now consider the specific case of systems with a second conservation law, that linearises such that the linear quantity has derivative zero. This commonly occurs in fluid systems for example, where the potential vorticity, and related quantities often have this form. With this extra condition we can make very general statements about the full asymptotic expansion.

\subsection{Enstrophy expansion}
Although not limited to fluid systems we will refer to the second conserved integral quantity as the enstrophy for clarity in this chapter. We start by expanding the enstrophy to 4th order in amplitude:
\begin{align} \nonumber
\mathcal{Z}=\mathcal{Z}^{(2)}+\mathcal{Z}^{(3)}+...&=\sum_p Z_p | a_p|^2+\sum_{pqr}Z_{pqr} a^*_p a^*_q a^*_re^{i\Omega_{pqr}t}\\
&+\sum_{pqrs}Z_{pqrs} a^*_p a^*_q a^*_s a^*_re^{i\Omega_{pqrs}t}+...
\end{align}
We then take derivatives in time:
\begin{align} \nonumber
\frac{\partial\mathcal{Z}}{\partial t}=\frac{\partial\mathcal{Z}^{(2)}}{\partial t}+\frac{\partial \mathcal{Z}^{(3)}}{\partial t}+...&=\sum_{pqr}Z_{p}C_{p}^{qr} a^*_p a^*_q a^*_re^{i\Omega_{pqr}t}+\sum_{pqr}i\Omega_{pqr}Z_{pqr} a^*_p a^*_q a^*_re^{i\Omega_{pqr}t}\\
&+\sum_{pqrs}C_{m}^{rs}Z_{pqm} a^*_p a^*_q a^*_s a^*_re^{i\Omega_{pqrs}t}+\sum_{pqrs}i\Omega_{pqrs}Z_{pqrs} a^*_p a^*_q a^*_s a^*_re^{i\Omega_{pqrs}t}+...
\end{align}
Initially we are interested in the lowest order of expansion:
\begin{align}
\frac{\partial\mathcal{Z}^{(2)}}{\partial t}=\sum_{pqr}Z_{p}C_{p}^{qr} a^*_p a^*_q a^*_re^{i\Omega_{pqr}t}+\sum_{pqr}i\Omega_{pqr}Z_{pqr} a^*_p a^*_q a^*_re^{i\Omega_{pqr}t}
\end{align}
Assuming that a triad can be isolated (discussion of when this is possible can be found in \cite{owen2018fast}) we choose one and label the modes $(1,2,3)$:
\begin{align}
(Z_{1}C_{1}^{23}+Z_{2}C_{2}^{31}+Z_{3}C_{3}^{12}) a^*_1 a^*_2 a^*_3e^{i\Omega_{123}t}+3i\Omega_{123}Z_{123} a^*_1 a^*_2 a^*_3e^{i\Omega_{123}t} \label{cons_of_enstrophy_trick}
\end{align}
Assuming exact triad resonances we have $\Omega =0$. We now use conservation of enstrophy to set this equal to zero and also assume exact resonances such that $\Omega_{123}=0$:
\begin{align}
Z_{1}C_{1}^{23}+Z_{2}C_{2}^{31}+Z_{3}C_{3}^{12} =0
\end{align}
Now where we have fast modes $Z_a=0$ and so for a triad where mode 1 is a zero mode and modes 2 and 3 are fast we get:
\begin{align}
Z_{1}C_{1}^{23} =0
\end{align}
and so the interaction coefficient is zero.\\
\\
We then continue to the next order. We choose to consider only the terms that are 4th order in amplitude:
\begin{align}
\frac{\partial \mathcal{Z}^{(3)}}{\partial t}+...&=\sum_{pqrs}(C_{m}^{rs}Z_{pqm}+i\Omega_{pqrs}Z_{pqrs}) a^*_p a^*_q a^*_s a^*_re^{i\Omega_{pqrs}t}
\end{align}

If we consider an isolated quartet (this is not in fact possible as Stokes\rq{} waves will always exist amongst the complex conjugates, however they only act to change the phase and so the amplitudes don\rq{}t change in size), and simultaneously demand that we have timescale separation (this was already done when we assumed equality with 0) then we can set $\Omega_{pqrs}=0$. We are left with the following:
\begin{align}
\frac{\partial \mathcal{Z}^{(3)}}{\partial t}=C_{a}^{12}Z_{34a}+C_{b}^{23}Z_{41b}+C_{c}^{34}Z_{12c}+C_{d}^{41}Z_{23d} \label{quad_cons}
\end{align}
Here a, b, c, d are modes such that the wavenumber is $\bm{k}_a=\bm{k}_1+\bm{k}_2$ etc.

We now make a brief aside to to discuss the form of the enstrophy we have expanded. Expressing the potential vorticity as $q$ and splitting it into linear and perturbation parts $Q$ and $q'$ we can write the generic form of enstrophy as: 
\begin{align}
Z=\frac{1}{2}q^2 = \frac{1}{2}(Q+q')^2.
\end{align}
The fast modes have $Q=0$, and so in general the cubic part of the expansion is limited in size to $O(\epsilon)$ for fast modes, due to the size of the $q'$ terms. In certain cases, for example the n-layer rotating shallow water equations (see \cite{owen2018fast} for the two layer case), the perturbation part takes the form $q'=Qf(\bm{x},t)$ for some function $f$. In this case if there are 2 fast modes in a triad the cubic part is identically zero. This removes the requirement used on equation \ref{cons_of_enstrophy_trick} of exact resonances, we can now assume that this is true for all triads, resonant or not. We can conclude that in general the fast fast slow interaction coefficient is always $O(\epsilon)$, and there are special cases that occur in well known equations such that they are identically 0.\\
\\
We now continue to the quartet order enstrophy expansion where we can use this to bound certain interaction coefficients. We see that if two or three of the modes in the $Z_{123}$ are fast modes such that $Q_1=Q_2=0$ for example then $Z_{123}=O(\epsilon)$. We now consider equation \ref{quad_cons} for modes 2, 3, 4 fast and mode 1 slow. We first consider the 1st and 4th terms. If modes a and d are fast or slow then we have the situation described above and $Z_{34a}=Z_{23d}=O(\epsilon)$. For the 2nd and 3rd terms this occurs for b, and c fast. However if b, and c are slow we can then consider the interaction coefficients. However from the previous order for fast modes interacting to form a slow we have: $C_{b}^{23}=C_{c}^{34}=0$. And so enstrophy is conserved to 3rd order. This will repeat almost exactly at every order. \\
\\
This method shows that conservation of enstrophy in interactions of many fast modes with a slow mode is entirely fulfilled by the triad interaction coefficient $C_1^{23}=0$. It is also irrelevant to this construction that the detuning needs to be zero as the $\mathcal{Z}^{(n)}$ term is always 0 for less than 2 slow mode combinations, although from elsewhere the scale separation will enforce this.\\
\\
\subsection{Direct proof by induction}
We can directly prove the weakness of the fast-fast-slow interactions using the generalised interaction coefficient derived in section \ref{ntet_exp}. This proof is valid to any order of the expansion, extending the conclusion of the previous section, and demonstrating the strength of using the generalised form of the interaction coefficients.\\
\\
We use the relation $C_{\bm{k}_1\bm{k}_2\bm{k}}^{ \pm\ \mp\ 0}=O(\epsilon)$ to extend to interactions of fast modes outputting a slow at all orders. By considering the nth order expansion we find that the effect of a set of only gravity waves on the slow mode is as follows:
\begin{align} \nonumber
\{C^{(n)}\}_{\bm{k}_1...\bm{k}_{n+1}\bm{k}}^{\pm...\ \pm\ 0}=&\\ \nonumber
\frac{1}{(n+1)!}\sum_{r=1}^{n}&\Bigg[ \sum_{\alpha_a,\alpha_b}\frac{ \{C^{(1)}\}_{\bm{k}_a\bm{k}_b\bm{k}}^{\alpha_a \alpha_b 0}\left. \{C^{(r-1)}\}_{\bm{k}_{n-r+2}...\bm{k}_{n+1}\bm{k}_b}^{\pm...\pm \alpha_b}\right|_{nr}\left. \{C^{(n-r)}\}_{\bm{k}_1...\bm{k}_{n-r+1}\bm{k_a}}^{\pm... \pm \alpha_a}\right|_{nr}}{-\{\Omega^{(n-r)}\}^{a}_{1,...,n-r+1}\{\Omega^{(r-1)}\}^{b}_{n-r+2,...,n+1}}\\&\qquad\qquad\qquad\qquad\qquad\qquad\qquad\qquad\qquad\qquad+ \textit{input wavenumber permutations}\Bigg]
\end{align}
We wish to use an inductive argument. We assume that $\{C^{(p)}\}_{\bm{k}_1...\bm{k}_{p+1}\bm{k}_a}^{\pm... \pm\  0}=O(\epsilon)$ for all $p<n$. The contribution for $\alpha_a=0$ will cause the right hand coefficient ($C^{(n-r)}$) to be $O(\epsilon)$. Similarly $\alpha_b=0$ will cause the middle coefficient ($C^{(r-1)}$) to be $O(\epsilon)$. But for the only remaining option, $\alpha_a,\alpha_b=\pm$, the lefthand interaction coefficient ($C^{(1)}$) has the same form and so that is also $O(\epsilon)$. We note also that all permutations have the same form in this case. Hence we have an inductive step, and can conclude that because a set of only fast modes cannot affect slow modes at first order($C^{\pm\pm 0}=O(\epsilon)$), this is true for all orders. The effect of fast modes on the slow will always be demoted in the asymptotic expansion to the next order. In cases where $\{C^{(p)}\}_{\bm{k}_1...\bm{k}_{p+1}\bm{k}_a}^{\pm... \pm\  0}=0$ (as occur in the n-layer shallow water equations for example, \cite{owen2019resonant}), these interactions are entirely absent from the expansion.\\
\\
By way of example we provide some of the specific implications in the case of the rotating shallow water equations. However much of the following is true for any conservative system with quadratic nonlinearity and a second conserved quadratic quantity such that the system supports zero modes eg stable stratification models and N-layer shallow water models.

\subsection{Rotating shallow water equations}
We will use the example of the rotating shallow water equations to demonstrate some of the corollaries of these expansions for zero mode systems. For this system there are three mode types: zero modes ($0$) and two fast modes ($\pm$). It can be shown (see \cite{embid1996averaging} for example) that the specific mode combination $(\pm, \mp, 0)$ has nonlinear interaction coefficient $C_{\bm{k}_1\bm{k}_2\bm{k}}^{\pm\ \mp 0}=0$. This is a consequence of the conservation of enstrophy to quadratic order, as shown in the previous section. It can also be shown from the form of the dispersion relation that no resonances are possible between three fast modes, or two slow and a fast mode.\\
\\
We consider a specific resonance (that studied in \cite{thomas2016resonant}):
\begin{align} \nonumber
Q_{\bm{k}_1\bm{k}_2\bm{k}_3\bm{k}}^{ 0\ \pm\ \mp\ 0}=\frac{2}{3}\left[ 
\sum_{\alpha_a}\frac{C_{\bm{k}_a\bm{k}_3\bm{k}}^{\alpha_a  \mp\  0}\left. C_{\bm{k}_1\bm{k}_2\bm{k_a}}^{ 0\   \pm\  \alpha_a}\right|_{nr}}{i(\omega_a-\omega_1-\omega_2)}
+\sum_{\alpha_b}\right.&\frac{C_{\bm{k}_b\bm{k}_1\bm{k}}^{\alpha_b  0\  0}\left. C_{\bm{k}_2\bm{k}_3\bm{k_b}}^{ \pm\   \mp\  \alpha_b}\right|_{nr}}{i(\omega_b-\omega_2-\omega_3)}  \left.
+\sum_{\alpha_c}\frac{C_{\bm{k}_c\bm{k}_2\bm{k}}^{\alpha_c  \pm\  0}\left. C_{\bm{k}_3\bm{k}_1\bm{k_c}}^{ \mp\   0\  \alpha_c}\right|_{nr}}{i(\omega_c-\omega_3-\omega_1)}\right]
\end{align}
Now using the resonance conditions and the fact that $C_{\bm{k}_1\bm{k}_2\bm{k}}^{ \pm\ \mp\ 0}=0$ we can reduce the possible combinations to $\alpha_a=0$, $\alpha_b=\pm$, and $\alpha_c=0$:
\begin{align} \nonumber
Q_{\bm{k}_1\bm{k}_2\bm{k}_3\bm{k}}^{ 0\ \pm\ \mp\ 0}=\frac{2}{3}\left[ 
\frac{C_{\bm{k}_a\bm{k}_3\bm{k}}^{0  \mp\  0}\left. C_{\bm{k}_1\bm{k}_2\bm{k_a}}^{ 0\   \pm\  0}\right|_{nr}}{i(\omega_a-\omega_1-\omega_2)}
+\sum_{\alpha_b=\pm}\right.&\frac{C_{\bm{k}_b\bm{k}_1\bm{k}}^{\alpha_b  0\  0}\left. C_{\bm{k}_2\bm{k}_3\bm{k_b}}^{ \pm\   \mp\  \alpha_b}\right|_{nr}}{i(\omega_b-\omega_2-\omega_3)}  \left.
+\frac{C_{\bm{k}_c\bm{k}_2\bm{k}}^{0  \pm\  0}\left. C_{\bm{k}_3\bm{k}_1\bm{k_c}}^{ \mp\   0\  0}\right|_{nr}}{i(\omega_c-\omega_3-\omega_1)}\right]
\end{align}
And so our quartets include combinations of slow-fast-slow triads and fast-fast-fast triads: the two interactions that did not occur to first approximation due to lack of resonances. We can therefore expect that any physical behaviour of the full system that these triads cause by moving energy between these modes is captured at second order in the asymptotic expansion, as a two stage movement of energy between modes.\\
\\
From the previous section we can also deduce that because $C_{\bm{k}_1\bm{k}_2\bm{k}}^{\pm\ \mp 0}=0$ all coefficients of the form $\{C^{(n)}\}_{\bm{k}_1 ... \bm{k}_{n+1}\bm{k}}^{\pm ... \mp 0}=0$.\\
\\
So we cannot have a resonant interaction between a single fast mode and the slow modes as this can never be resonant, and we cannot have an interaction between a single slow mode and the fast modes as this will always have 0interaction coefficient. This limits our interactions with a zero mode output so that for the $n$th closure with resonance between $n+2$ modes the number of fast modes ($n_f \in \mathbb{N}$) must be in the range $1< n_f< n+1$ (hence at first closure there is no possible value of $n_f$). This is our generalisation of the conclusion in \cite{thomas2016resonant} that there exist energy exchanges between fast and slow modes at closures above first order. We can in fact go further. The general theory is that the first order of expansion is dominant (for example: for RSWE this is triad order, for surface waves quartet order) however for systems with distinguished zero modes there will always be interactions that are never present at triad order, and hence the full physics is not in the model unless both triad and quartets are considered. We suggest that above this order there is no benefit to greater expansion, as all physics is included and higher orders will simply be weaker interactions important only on longer timescales.\\
\\
In addition we know that in the absence of fast modes there cannot be interactions between slow modes from anything other than triads, as every triad is resonant.  However when there are fast modes present (even just at $O(\epsilon)$ in the form of slaved \lq{}fast-like\rq{} modes) there is a quartet `pathway' via slaved modes of wavenumbers $\bm{k}_a$, $\bm{k}_b$, $\bm{k}_c$ as defined in the definition (\ref{Qdef}). This gives rise to the \lq{}higher order quasi-geostrophic\rq{} expansions of \cite{zeitlin2003nonlinear} and \cite{thomas2016resonant}. Of particular note is the difference between the case in the infinite domain (with compact ICs) and the case in the periodic domain. Both authors find that in the infinite domain the four slow mode interactions occur but that the two fast two slow do not. The explanation for the absence of the second type was propagation of the fast modes out of a shared domain, removing the interactions. However the fast-like slaved modes of the first type actually have a zero group velocity and so do not leave the domain of interest, allowing this interaction to happen in the continuous domain as well as the periodic.\\
\\
We now compare the near-resonant expansion to our conclusion that all \lq{}physics\rq{} exists at quartet order in the exact resonant expansion. Our description of the near-resonances picking out the most significant of the behaviour from the exact expansion at all orders suggests that we would expect the same effect: all possibly interactions will necessarily be present, and hence all physics. We can compare this conclusion to that of several numerical studies such as \cite{smith2005near} and \cite{smith1999transfer} that shows near-resonant numerical simulations can introduce the qualitative effects of the full equations. For this conclusion we do need to extend the notion of zero-mode systems to that of $\epsilon$-mode systems - the introduction of the $\beta$-plane assumption breaks the symmetry that leads to zero-modes, however we can make the assumption that these modes are $\epsilon$ close to the zero modes (a similar assumption is made in \cite{buhler1998non}) and treat them as a small perturbation from the zero-mode systems we have described here.

\section{Discussion}
\label{discussion}
In this paper we have examined quadratic and higher order wave interactions in a general class of nonlinear wave equations. with quadratic nonlinearity, using a multi-time scale asymptotic expansion, We have found a general form of the interaction coefficient valid for to all orders of the expansion, This general form is used to demonstrate how including the near resonant interactions at the triad order is equivalent to picking out the set of the fastest higher order interactions, distinguished by the size of the interaction coefficient. We then showed that in the presence of a second conserved quantity, of a form common to geophysical fluid systems that preserve some analogue of vorticity, that there then are whole families of interactions between modes that are bounded to $O(\epsilon)$ or smaller, effectively demoting them to a slower timescale.\\
\\
Although our motivation has come for geophysical dynamics, the present results are widely applicable to any physical system of the form (\ref{nl_eq_int1}) with a constraint of the form (\ref{Qconstraint2}). Here we note that the restriction to quadratic nonlinearity is not severe, as systems with cubic or higher order nonlinearity can sometimes be reduced to systems with just quadratic nonlinearity by changing the dependent variable, or introducing further dependent variables. generalised to nonlinear wave systems with all orders of nonlinearities. However, nonlinearity of the form encountered for water waves for instance presents more of a challenge and to remains to be seen if the present approach can be extended to such systems.\\
\\
This work, like many others, has assumed that at the linearised order the waves are periodic with slowly-varying time-dependent amplitudes. However, if slow-variation is the spatial variables is also permitted leading to the notion of wave packets and group velocity then we might expect rather different behaviour, see \cite{reznik2001nonlinear} and\cite{owen2019resonant} for instance. This would seem to be a fruitful area for further research.

\clearpage
\printbibliography

\end{document}